\documentclass[prb,twocolumn,aps,superscriptaddress]{revtex4-2}

\usepackage{graphicx}
\usepackage{dcolumn}
\usepackage{bm}
\usepackage{amssymb}
\usepackage{amsmath}
\usepackage{physics}
\usepackage{sublabel}
\usepackage[utf8]{inputenc}
\usepackage{hyperref}
\usepackage[usenames, dvipsnames]{color}
\usepackage[english]{babel}
\usepackage[T1]{fontenc}
\usepackage{bbm}

\usepackage{float}
\usepackage{verbatim}
\hypersetup{colorlinks = true, linkcolor=blue, citecolor=blue, urlcolor=blue}

  \makeatletter
    \renewcommand\@make@capt@title[2]{%
     \@ifx@empty\float@link{\@firstofone}{\expandafter\href\expandafter{\float@link}}%
      {\textsc{#1}}\@caption@fignum@sep#2\quad}%
    \makeatother

\begin{document}


\title{Proposal for a solid-state magnetoresistive Larmor quantum clock}


\author{Amal Mathew}
\affiliation{Department of Physics, Indian Institute of Technology Bombay, Powai, Mumbai-400076, India}
\author{Kerem Y Camsari}
\affiliation{Department of Electrical and Computer Engineering, University of California, Santa Barbara, Santa Barbara, CA 93106, USA}
\author{Bhaskaran Muralidharan}
\affiliation{Department of Electrical Engineering, Indian Institute of Technology Bombay, Powai, Mumbai-400076, India}
\email{bm@ee.iitb.ac.in}


\date{\today}

\begin{abstract}
We propose a solid-state implementation of the Larmor clock that exploits tunnel magnetoresistance to distill information on how long itinerant spins take to traverse a barrier embedded in it. Keeping in mind that the tunnelling time innately involves pristine pre-selection and post-selection, our proposal takes into account the detrimental aspects of multiple reflections by incorporating multiple contacts, multiple current measurements and suitably defined magnetoresistance signals. Our analysis provides a direct mapping between the magnetoresistance signals and the tunneling times and aligns well with the interpretation in terms of generalized quantum measurements and quantum weak values. By means of an engineered pre-selection in one of the ferromagnetic contacts, we also elucidate how one can make the measurement ``weak'' by minimizing the back-action, while keeping the tunneling time unchanged. We then analyze the resulting interpretations of the tunneling time and the measurement back action in the presence of phase breaking effects that are intrinsic to solid state systems. We unravel that while the time-keeping aspect of the Larmor clock is reasonably undeterred due to momentum and phase relaxation processes, it degrades significantly in the presence of spin-dephasing. We believe that the ideas presented here also open up a fructuous solid state platform to encompass emerging ideas in quantum technology such as quantum weak values and its applications, that are currently exclusive to quantum optics and cold atoms. 
\end{abstract}

\maketitle

\begin{figure}[h]
    \centering
    \includegraphics[width=1.058\linewidth]{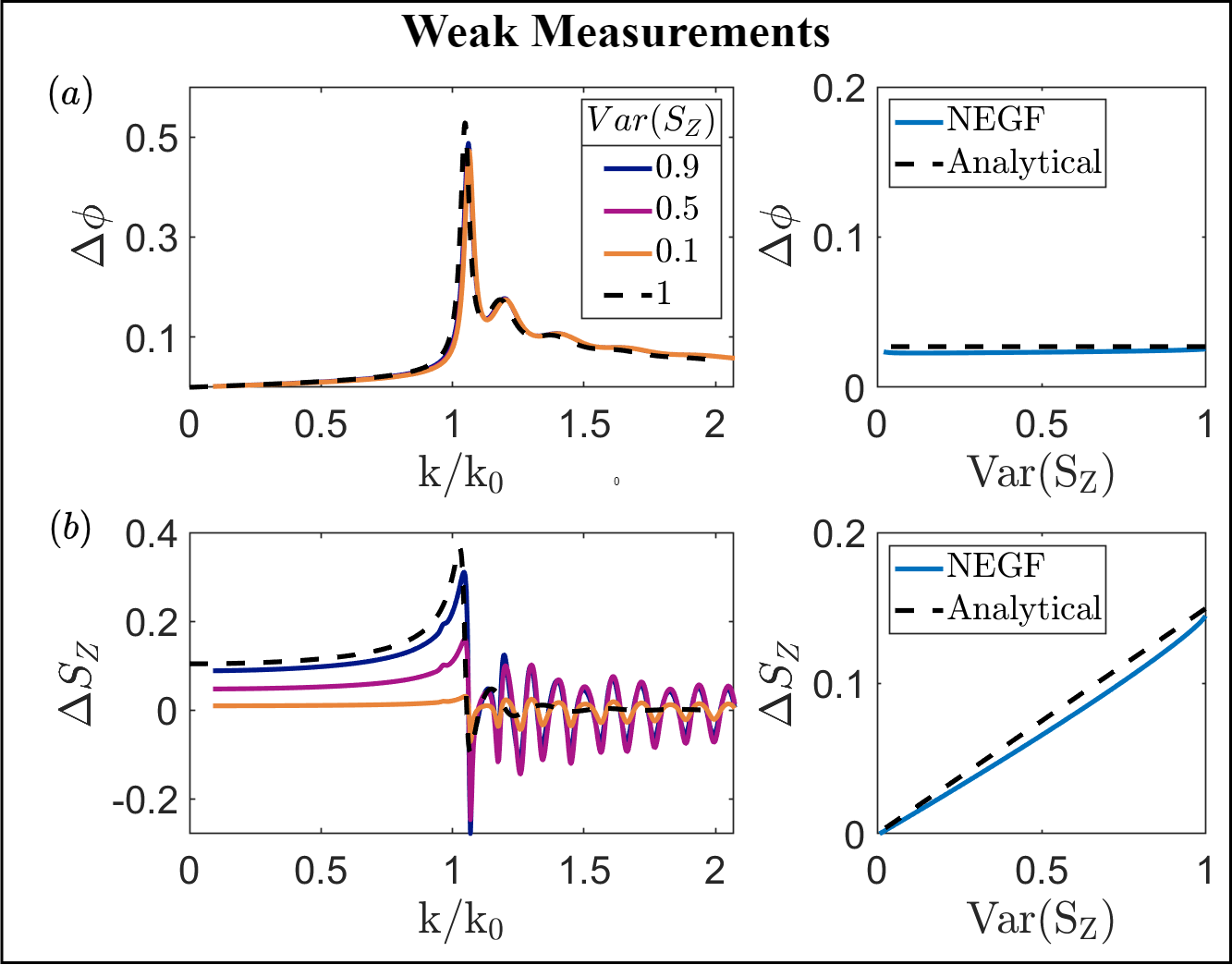}
    \caption{Minimizing the measurement back action. (a) Precession angle of electron spin about the $\hat{z}$ axis ($\Delta \phi$) is found to remain constant whereas (b) the measurement back-action, $\Delta S_Z$ decreases proportional to the variance in $S_Z$,  indicating that the pointer deflection is unaffected. This aligns well with Ref.\cite{Steinberg_1995} \ (see the corresponding right insets for results at $k=0.8k_0$), that the tunnel time is indeed $\tau_Y$, which remains constant as the measurement is made weaker.  }
    \label{fig:weak}
\end{figure}

Despite the lack of a  ``time operator'' in quantum mechanics \cite{Pauli_Objection}, quantum time keeping can be connected with the measurement of space-time distances \cite{Wigner,Peres,choi2013} signifying the passage of time needed for a quantum process to occur, most generally, between two spatial co-ordinates \cite{Wigner, Peres}.  The tunneling time - the time a particle takes to tunnel through a barrier, that has been a subject at the heart of hot debates in physics \cite{Landauer_RMP_1994,Winful_PRL,WINFUL20061,KOFMAN201243,Satya_Sainadh_2020} precisely fits into this paradigm.  B\"uttiker \cite{buttikerlandauer1982,buttiker1983}, following earlier works \cite{Baz, Baz2, Rybachenko}, solidified a construct - the Larmor clock to estimate the tunneling time, which is based on the Larmor precession of a stream of spins inside a barrier subject to a weak Zeeman field perpendicular to the plane of the precession. This idea was further elegantly interpreted in the perspective of generalized von Neumann measurements \cite{Steinberg_1995}, with the tunneling time proportional to a quantum weak value \cite{Vaidman_1,Sudarshan,Vaidman_2,Boaz,Jordan_PRL,Jordan_RMP,choi2013}. A holistic viewpoint of the tunnel time problem requires delving into the following intertwined aspects: a) the construct of the Larmor clock that is based on a straightforward analysis of spin dependent tunneling \cite{buttiker1983} and the description of the tunneling time and the dwell time from this analysis, b) its connection with generalized von Neumann measurements in relation to a generic description of quantum time keeping, and c) that the pre-selection and post-selection of quantum states are inherently involved which necessitates a connection to quantum weak values \cite{Steinberg_1995}.  \\
\indent Recent ground breaking experiments on this topic using a cold atoms realization of the Larmor clock \cite{Ramos2020,PhysRevLett_Steinberg} open the possibility of making the time keeping aspects as well as the aspects related to quantum weak values accessible to a larger class of experiments.
\begin{figure*}
    \includegraphics[width=0.9\textwidth]{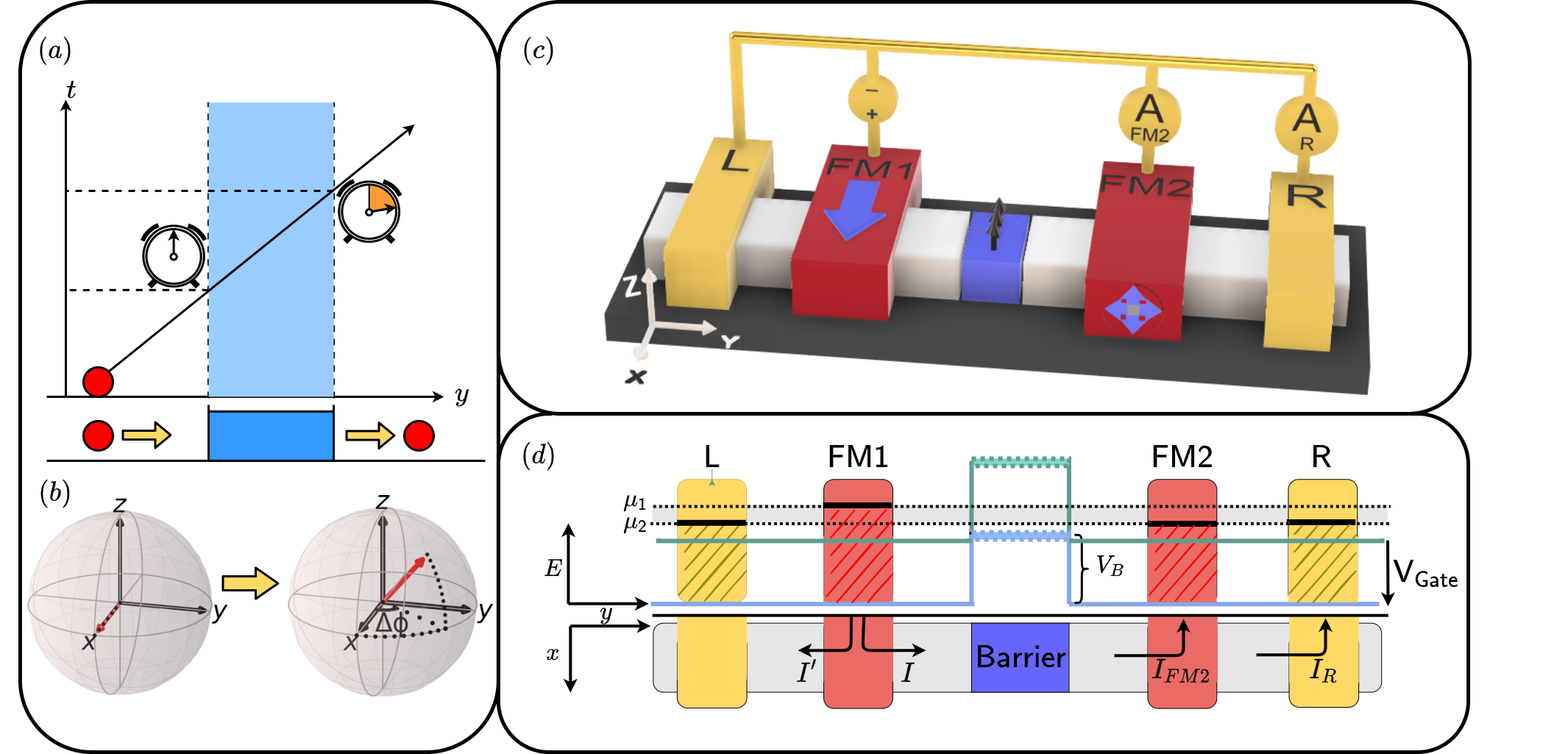}
    \caption{Schematics: (a) Depiction of quantum time keeping in terms of tunneling time estimation. A pointer movement attached with the particle tracks the time taken. (b) A schematic of Larmor precession as well as the alignment in $\hat{z}$ direction of a tunnelling electron. (c) The proposed magnetoresistive setup. The yellow contacts are NM contacts whereas the red contacts are FM in the direction specified by the blue arrows. The current measurements across the two ammeters are used to deduce the necessary transport signals. Schematic along the transport $\hat{y}$ direction (d) Because tunnelling time is defined for a particular value of $k$, we consider a low-bias situation in which only electrons within a small energy range conduct current. The corresponding $k$ values can be modulated by using a gate voltage to scan through the $k$ space}
    \label{fig:1}
\end{figure*}
Given the current progress in nanoelectronics, and especially  nanomagnetism and spintronics \cite{Zutic,FERT2019817,Hirohita}, a suitably designed solid state device platform can potentially provide for an indispensable test bed to integrate such emerging ideas into such a platform. The object of this Letter is to propose a prototype solid state spintronic test bed that caters to the holistic viewpoint of quantum time keeping described above. We also believe that the ideas presented here can encompass emerging ideas in quantum technology such as quantum weak values and its applications that are currently exclusive to quantum optics and cold atoms \cite{Jordan_RMP}.  \\
\indent  Before delving into our setup, we briefly describe the generics of quantum time keeping in connection with the Larmor clock, for which, we refer to Fig.~\ref{fig:1}(a) and (b). The measurement of tunnel time can be thought of in terms of a pointer that gets ``kicked'' as the particle tunnels through the barrier. The difference between the initial and final pointer readings can be used to decipher the time taken for the process.
In the Larmor clock, as depicted in Fig.~\ref{fig:1}(b), the spin orientation of the particle along the $x-y$ plane acts as the pointer, and the in-plane angle of rotation denotes the pointer reading. Based on this, for the in-plane rotation to act as a viable pointer it becomes crucial to have a well defined pre-selection and post-selection \cite{Steinberg_1995,choi2013} of states, at the incident and at the transmitted regions respectively. \\
\indent Our proposal is schematized in Fig.~\ref{fig:1}(c), in which we utilize tunnel magnetoresistance transport signals to distill the necessary information on the tunneling time of itinerant spins traversing a barrier embedded in it. Ferromangetic contacts take into account the crucial aspects involving the pre-selection and the post-selective measurement. Unlike the cold atoms implementation however, spintronic devices suffer from setbacks such as multiple reflections at the ferromagnetic contacts, impurities and channel phase breaking processes. These aspects are serious impediments specifically to pre-selection and post-selection of the states that the proposal heavily relies on. The detrimental aspect of multiple reflections, we show, can be mitigated using additional ``padding contacts'' and incorporating multiple current measurements that will be described in detail. \\
\indent Using the Keldysh non-equilibrium Green's function (NEGF) technique \cite{Datta,Meir-Wingreen-1992,haug2007quantum} to calculate the transport signals, we first demonstrate that our analysis provides a direct mapping between the magnetoresistance signals and the Larmor tunneling times. Our results also consistently align with the interpretation of the tunneling time as a quantum weak value, with the real and imaginary parts signifying the tunneling time and the measurement back action \cite{Steinberg_1995,Jordan_RMP}. By means of an engineered pre-selection in one of the ferromagnetic contacts, we further elucidate how one can make the measurement ``weak'' by minimizing the back-action, while keeping the tunneling time unchanged. \\
\indent We further analyze the resulting interpretations of the tunneling time and the measurement back action in the presence of phase breaking effects \cite{DANIELEWICZ1984239,Datta,PhysRevB.75.081301,7571106,doi:10.1063/1.5023159,doi:10.1063/1.5044254,PhysRevApplied.8.064014,camsari2020,Duse_2021}, that are intrinsic to solid state systems. We uncover that, while the time-keeping aspect of the Larmor clock is reasonably undeterred due to momentum and phase relaxation processes, it degrades significantly in the presence of spin-dephasing. We now formalize the three crucial aspects that were described earlier, before describing the setup in detail and the results to follow. \\
{\it{The Larmor clock:}} For the orientation we consider, $\hat{x}$-polarized spins tunnel through a barrier that encloses a Zeeman field in the $\hat{z}$-direction. Inside this barrier the spins undergo Larmor precession in the $x-y$ plane and a damping that tries to orient the spins along the $\hat{z}$-direction. In the weak magnetic field limit, the orientation of the average spin $\langle S \rangle$ of the outgoing stream dictates the tunnel time, which can be written as 
\begin{equation}
\begin{aligned}
\left\langle S_{Z}\right\rangle&=(\hbar / 2) \omega_{L} \tau_{Z} \\
\left\langle S_{Y}\right\rangle&=-(\hbar / 2) \omega_{L} \tau_{Y} \\
\left\langle S_{X}\right\rangle&=(\hbar / 2)\left(1-\omega_{L}^{2} \tau_{X}^{2} / 2\right),
\end{aligned}
\label{eqn:1}
\end{equation}
where $\hbar\omega_L/2$ is the Zeeman energy and $\omega_L$ is the Larmor frequency. Although $\tau_Y$ and $\tau_Z$ are purely mathematical constructs that describe various times involved in the tunneling process, B{\"u}ttiker argued that the actual tunneling time is given by $\tau_T = \sqrt{\tau_Y^2+\tau_Z^2}$. It was further remarked that $\tau_T$ is the tunneling traversal time and that in the case of symmetric barriers like the ones considered here, it is equal to another quantity called the dwell time $\tau_d$. Keeping in mind various stimulating discussions in this field, we will follow the interpretation based on generalized measurements \cite{Steinberg_1995}.\\
{\it{Generalized measurements:}} To formalize the above discussion, we refer back to the schematic in Fig.~\ref{fig:1}(a), which depicts space-time measurement. Connecting with the theory of generalized measurements discussed in the supplementary section, the $\hat{S}_z$ operator is the generator of the Larmor precession angle $\hat{\phi}$ for measuring the spatial barrier operator $\hat{U}(y)$. The interaction Hamiltonian that defines the ``measurement'' process is then given by $\hat{H}_{int} = g \mu_B B_z \hat{S}_z \hat{U}(\hat{y})$, which represents the standard Zeeman interaction Hamiltonian with a magnetic field $B_z$ along the $\hat{z}$ direction, but only limited to the barrier region represented by the barrier function $\hat{V}( \hat{y})$. Connecting with the B\"uttiker clock, the stream of $\hat{x}$-polarized electrons form the pre-selection and the measurement along $\hat{y}$ or $\hat{z}$ direction forms the post-selection process. \\
{\it{Quantum weak values:}} While the treatment of the Larmor clock gives a straightforward prescription for calculating $\tau_{Y(Z)}$, it can be established that $\tau_Y$ and $\tau_Z$ from \eqref{eqn:1} actually translates to the real and imaginary parts of the weak value of the measurement process described above, and is defined as 
\begin{eqnarray}
\tau_Y =\frac{m}{\hbar k } \bf{Re}\left ( \frac{\langle f \mid \hat{U}(\hat{y}) \mid  i \rangle}{ \langle f \mid i \rangle} \right) \\ \nonumber
\tau_Z =\frac{m}{\hbar k } \bf{Im}\left ( \frac{\langle f \mid \hat{U}(\hat{y}) \mid  i \rangle}{ \langle f \mid i \rangle} \right), \\
\end{eqnarray}
where $ \langle y | i (f) \rangle$ represents the wave-function of the incident (transmitted) stream of spins. The incident and the transmitted beams represent the pre-selection and the post-selection respectively. The weak value has both real and imaginary components and thus in this interpretation, $\tau_Y$ alone is the tunneling time, whereas $\tau_Z$, the imaginary part represents the back action due to the measurement process.\\
{\it{Magnetoresistive Setup:}}  Consolidating the concepts discussed above and elaborated in the supplementary material, we now proceed to a detailed exposition of the magnetoresistive setup, shown in Fig.~\ref{fig:1}(c). The device region consists of a long enough single moded channel with a barrier in the middle where a small Zeeman field $B_z$ is applied along the $\hat{z}$-direction, with the Zeeman splitting energy $V_Z$. Two normal metallic (NM) contacts are placed at the ends in order to manipulate reflections and hence produce a viable transport signal at the ammeters. The ferromagnetic (FM1) contact on the left side injects the $\hat{x}$- polarized stream of electrons, and the ferromagnetic contact (FM2) on the right side is used as the detector for post-selective measurement, whose orientation is along the $\hat{y}$-direction or the $\hat{z}$-direction, so as to measure $\tau_Y$ or $\tau_Z$ respectively. The unpolarized contact to the left of FM1 acts as a sink that collects the reflected waves from the barrier. We read the currents through the $\pm \hat{y}$ (or $\pm \hat{z}$) polarized FM2 contacts which are also grounded. Since the current drains into these contacts, and they are located "downstream" from the barrier, post-selection rules are also satisfied and the measured current is composed of electrons that have tunnelled through the barrier. \\
\indent The entire setup is back-gated such that a gate voltage $V_{G}$ can add an energy offset to the entire channel, in order to select a particular carrier momentum $k$. Figure~\ref{fig:1}(d) shows the schematic of the cross section of the setup, with the two NM contacts kept at electrochemical potentials $\mu_1$ and $\mu_2$. For the transport measurement, a small electrochemical potential difference $\mu_1-\mu_2=eV$ is maintained such that a small applied voltage can inject the desired momentum $k$ for the incoming stream of electrons. \\ 
{\it{Transport Signals:}}  Following a detailed analysis (see supplementary material) of multiple reflections at the FM contacts, we can show that the transport signal $D_Y$ related to $\tau_Y$ can be derived based on the currents registered at the FM2 contact:
\begin{equation}
D_Y = \frac{I^{+}_{FM2} - I^{-}_{FM2}}{I^{+}_{FM2} + I^{-}_{FM2}},
\label{eqn:4}
\end{equation}
where $I^{+}_{FM2}$ and $I^{-}_{FM2}$ represent the measured currents at the FM2 contact when it is polarized in the $+\hat{y}$ and $-\hat{y}$ directions respectively. This transport signal $D_Y$ is a measure of polarization of $\hat{y}$-spin of electrons in the channel. Specifically, we have the average post-selected $\hat{y}$- component of the spin as:
\begin{equation}
\left<S_Y\right> = (\hbar/2) \frac{I^{+}_{FM2} - I^{-}_{FM2}}{I^{+}_{FM2} + I^{-}_{FM2}} = -(\hbar/2)\omega_L \tau_Y.
\label{eqn:5}    
\end{equation}
We can therefore write the tunneling time $\tau_Y$ in terms of currents observed at the FM2 contact as:
\begin{equation}
\tau_Y = -\frac{1}{\omega_L} \frac{I^{+}_{FM2} - I^{-}_{FM2}}{I^{+}_{FM2} + I^{-}_{FM2}}.
\label{eqn:6}    
\end{equation}
\begin{figure*}[t]
    \centering
    \includegraphics[width = 0.7\linewidth]{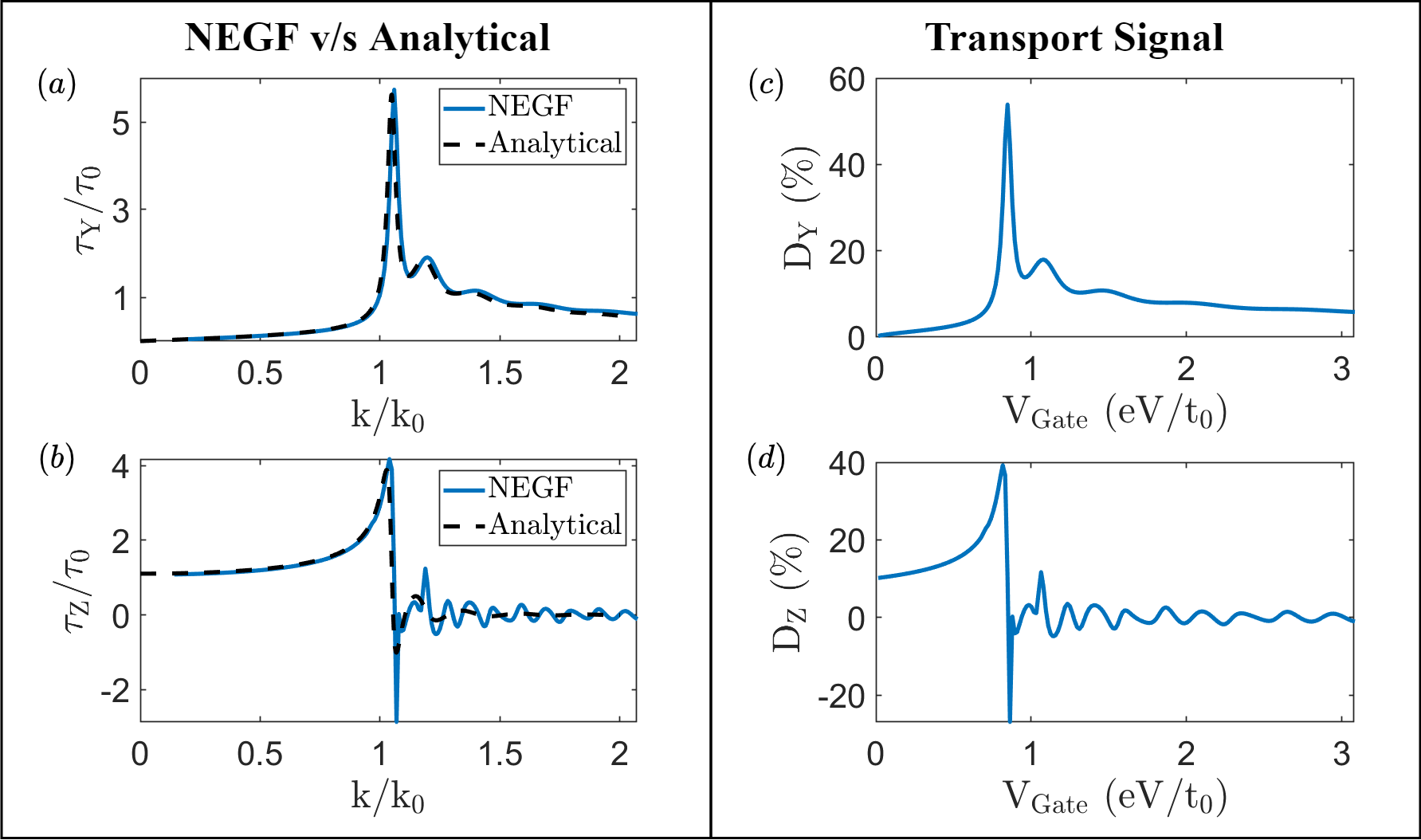}
    \caption{\footnotesize Transport signals and comparison with the B\"uttiker model \cite{buttiker1983}. (a) Tunneling time and (b) measurement back-action as a function of $k$ as predicted by NEGF model which matches the analytical results from the B\"uttiker model. (c) and (d) The corresponding magnetoresistive transport signal $D_Y$ and $D_Z$ respectively. For our setup, a twenty percent measurable polarization gives the embedded information on $\tau_Y$ and $\tau_Z$ as seen in  (a) and (b).}
    \label{fig:4}
\end{figure*}
The derivation of the transport signature $D_Z$ that captures the measurement back action $\tau_Z$, as explained in supplementary material is more involved and is given by:
\begin{equation}
D_Z= -\frac{(I_R^{+}-I_{FM2}^{+}) - (I_R^{-}-I_{FM2}^{-})}{(I_R^{+}-I_{FM2}^{+}) + (I_R^{-}-I_{FM2}^{-})},
\label{eqn:7}    
\end{equation}
where $I^{\pm}_{FM2}$ are, similar to the previous case, the currents through the FM2 contact while it is in the $\pm \hat{z}$ orientation respectively. $I_{R}^{\pm}$ are the currents measured in the second ammeter connected to the right NM contact when the ferromagnetic contact is in the $\pm \hat{z}$ orientation respectively. Thus, $D_Z$ is analogous to $D_Y$ defined in \eqref{eqn:4}, and correspondingly measures the post-selected $\hat{z}$-spin which is given by
\begin{equation}
\left<S_Z\right> = -(\hbar/2) \frac{(I_R^{+}-I_{FM2}^{+}) - (I_R^{-}-I_{FM2}^{-})}{(I_R^{+}-I_{FM2}^{+}) + (I_R^{-}-I_{FM2}^{-})} = (\hbar/2)\omega_L \tau_Z.
\label{eqn:8}    
\end{equation}
Thus, the measurement back-action in terms of contact currents is given by:
\begin{equation}
\tau_Z = -\frac{1}{\omega_L} \frac{(I_R^{+}-I_{FM2}^{+}) - (I_R^{-}-I_{FM2}^{-})}{(I_R^{+}-I_{FM2}^{+}) + (I_R^{-}-I_{FM2}^{-})}.
\label{eqn:9}    
\end{equation}
{\it{Coherent transport:}} With the above formulation, we evaluate the transport signal currents using the Keldysh NEGF technique \cite{Datta,Meir-Wingreen-1992} detailed in supplementary materials section. With the terminal current operator $\hat{I}^{\alpha}_{op}$, where $\alpha= FM1, FM2$, we can find quantities related to spin currents as $I^s_{\alpha}= {\bf{Tr}}\left[\hat{\sigma} \hat{I}_{\alpha}\right]$, where $\hat{\sigma}$ represents the vector Pauli spin operator. The channel is written in the tight-binding representation of the one-band effective mass Hamiltonian with an on-site energy $E_0$, and hopping energy $t_0$. The barrier region in the middle has a potential $V_B$, and is subject to a Zeeman energy $V_Z$ along the $\hat{z}$-direction. In order to resolve carrier momenta, we relate the energy of the carrier with the gate potential given by:
\begin{equation} 
E = -eV_G + 2t_0(1- \cos(ka)).
\label{eqn:10}
\end{equation}
Assuming $E= 0$ without loss of generality, such that the gate potential gives the required energy translation, gives the necessary transformation between carrier momenta and gate potential. 
 \\
\indent We observe that in the coherent ballistic regime, the results from our simulation are a near perfect match with the analytical results derived by B{\"u}ttiker as shown in Fig.~\ref{fig:4}. Figures \ref{fig:4}(c) and \ref{fig:4}(d) show the magnetoresistance signature $D_Y$ as a function of the gate voltage, thus correlating a magnetoresistance measurement with the Larmor tunnel time.  We clearly see how an experimental setup that tracks the transport signals $D_Y$ and $D_Z$ that can indeed yield the tunneling time, as well as a measure of the back action. This aspect constitutes the crux of our solid state Larmor clock.\\
\indent Having demonstrated the setup in terms of reproducing transport signals that connect to the free-space B\"uttiker proposal, we move on to analyze a few realistic effects. First, we see what happens with realistic barriers and then move on to effects that relate to dephasing that naturally occurs in such solid state setups.\\
{\it{Realistic barriers:}} Instead of a perfectly rectangular barrier, we now consider a barrier of the form
\begin{equation}
U(y) = V_B \left(\tanh(L/2-y)+\tanh(L/2+y)\right)/2.
\label{eqn:12}    
\end{equation}
This is implemented by replacing the perfect square barrier potential in the channel Hamiltonian by a the appropriate function defined above.
\begin{figure}[h]
    \hspace{-1.6em}
    \centering
    \includegraphics[width=1.049\linewidth]{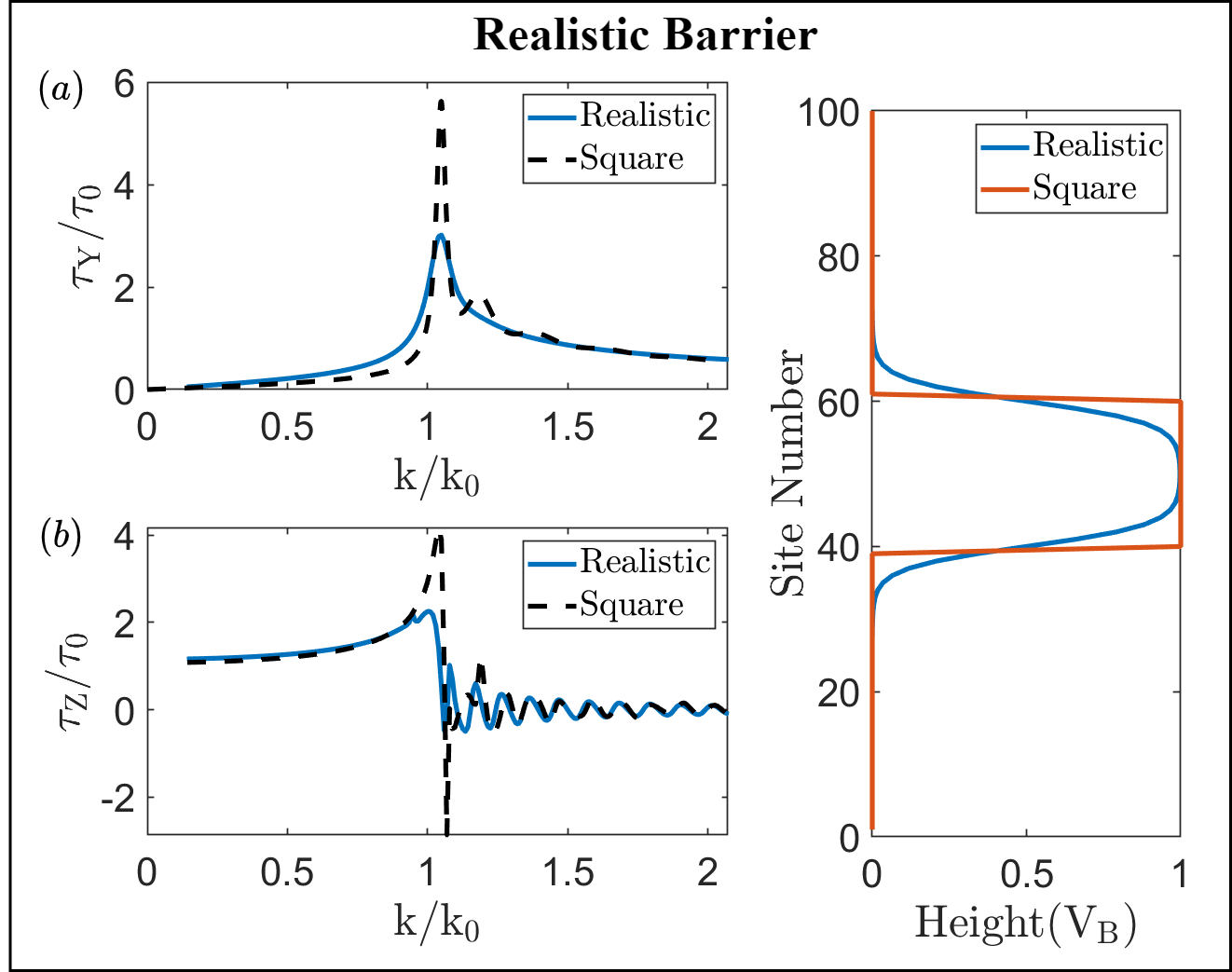}
    \caption{ Realistic barrier. (a) The tunnel time, and (b) measurement back action profile as a function of carrier momentum $k$ for a realistic barrier with band bending effects. We clearly note the effect of the barrier narrowing near the top of the barrier and widening near the bottom.}
    \label{fig:5}
\end{figure}
We find that for such a barrier, in the tunneling regime, i.e., $k/k_0 < 1$, where $k_0 = \sqrt{2mV_B}/\hbar$, electrons take longer time to traverse the barrier. However, the maxima of the tunneling time is smaller than that of the rectangular barrier. This is expected since the barrier has indeed thinned near the top as expected in a typical band-bending situation created upon contacting dissimilar materials. We also notice that the tunneling time is greater at smaller energies due to the widening of the barrier in those regions. \\
{\it{Minimizing the measurement back action :}}
We now focus on the interpretation of the imaginary part $\tau_Z$ that relates to the measurement back action. According to Steinberg \cite{Steinberg_1995}, it possible to make this measurement ``weaker'' and reduce the measurement back-action by preparing the electron in a spin-squeezed initial state that increases the uncertainty in the pointer position. Although squeezed states are impossible to prepare from a solid state perspective, we present an alternate method that can exhibit similar phenomena. We consider injected electrons with its spin oriented in the $x-z$ plane as opposed to the $\hat{x}$ orientation considered earlier. This decreases the uncertainty in $S_Z$ and increases the uncertainty in the pointer $\hat{\phi}$ position. We have from \cite{Steinberg_1995} that the change in pointer position (corresponding to $\tau_Y$) and the pointer momentum (corresponding to $\tau_Z$) is related to the uncertainty in pointer position as:
\begin{equation}
\begin{aligned}
    \Delta \phi &= \omega_L \tau_Y = k\Re \left<U(y)\right>_{fi}\\
    \Delta S_Z &= \omega_L \tau_Z = k\Im \left<U(y)\right>_{fi}/2\sigma^2,
\end{aligned}
\label{eqn:13}
\end{equation}
where $\left<U(y)\right>_{fi}$ is the weak value of the barrier function $U(y)$, $\sigma^2$ is the variance in the $\phi$ distribution. This implies that the precession angle remains constant whereas the measurement back-action decreases proportional to a decrease in uncertainty in $S_Z$. This is explicitly verified using our calculations on our setup, where on comparing Fig.~\ref{fig:weak}(a) and Fig~\ref{fig:weak}(b), we clearly notice the measurement back action decreasing as the variance in $S_Z$ decreases, while keeping the signal pointer position $\Delta \phi$ unchanged. This implies clearly that the measured tunnel time is indeed $\tau_Y$, thus complying with the interpretation in Ref.\cite{Steinberg_1995}.\\
\\
\begin{figure*}[t]
    \centering
    \includegraphics[width=0.85\linewidth]{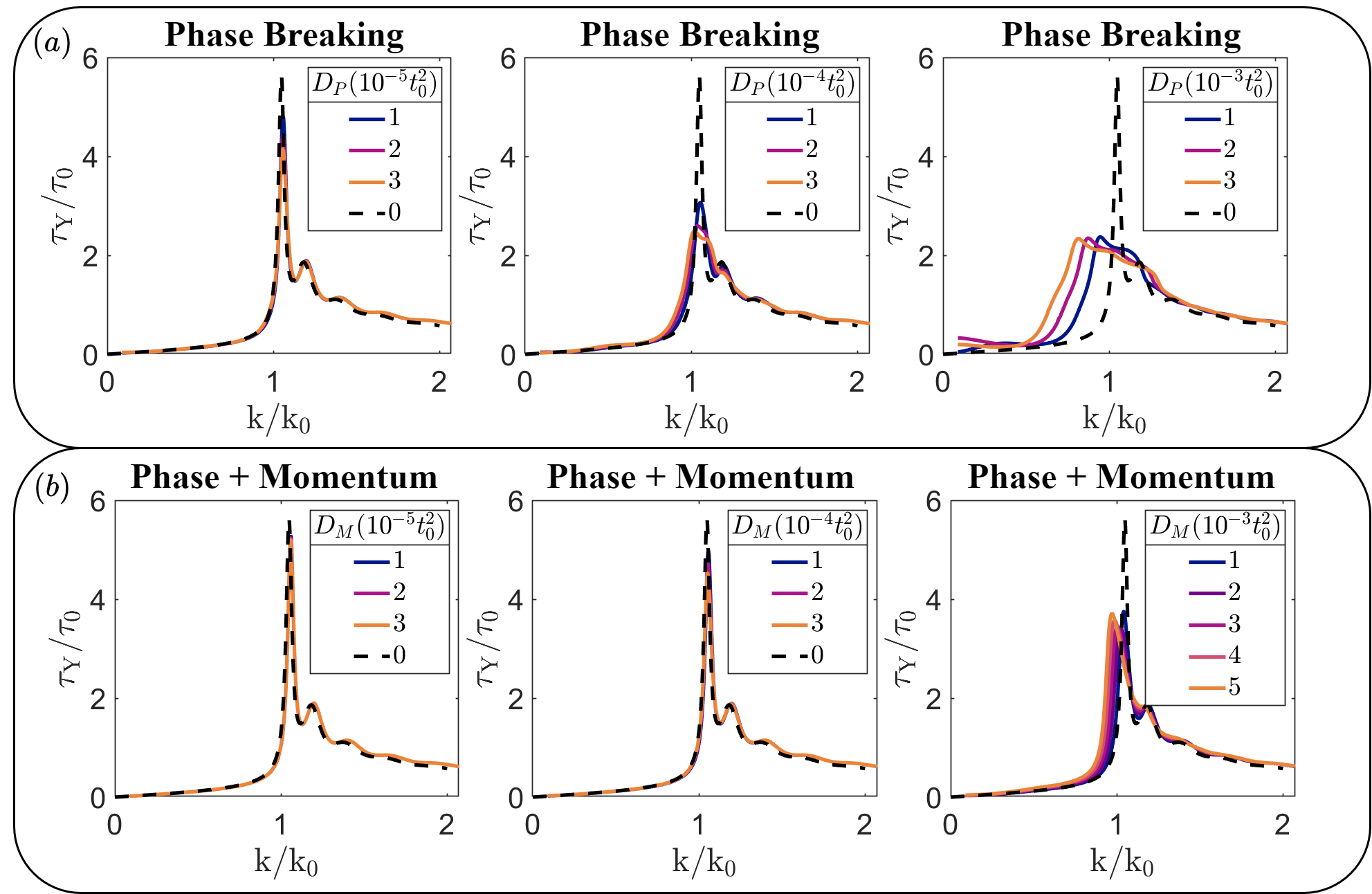}
    \caption{Effect of phase relaxation and momentum relaxation. Tunnel time $\tau_Y$ profiles for varying strengths of the dephasing interaction parameter for (a) pure phase relaxation, and (b) for momentum relaxation interactions. Note that the dephasing interactions in (a) and (b) are spin preserving and the observed disturbance in the transport signal is purely a result of disturbance in the tunneling phenomenon itself.}
    \label{fig:6}
\end{figure*}
{\it{Channels with dephasing:}} Dephasing interactions that are typical in solid state systems, typically give rise to phase breaking processes that would degrade the crucial phase coherent nature of the spins. Typical interactions of this kind include pure phase relaxation via electron-electron interactions, momentum and phase relaxation via fluctuating local non-magnetic impurities, and spin relaxation via magnetic impurities. These aspects can be added phenomenologically within the framework of Keldysh NEGF formalism \cite{DANIELEWICZ1984239,Datta,PhysRevB.75.081301,7571106,PhysRevApplied.8.064014,camsari2020,Duse_2021} via appropriate dephasing self-energies. \\
\indent Pure phase as well as momentum relaxation processes \cite{PhysRevB.98.125417} within the channel can be added via a scattering self-energy and its related in-scattering self-energy \cite{camsari2020} in its matrix form as
\begin{equation}
\begin{aligned}
\left[\boldsymbol{\Sigma}^{r}_{s}\right]_{ij}&=D_{ijkl} \left[\mathbf{G}^{r}\right]_{kl} \\
\left[\boldsymbol{\Sigma}^{<}_{s}\right]_{ij}&=D_{ijkl} \left[\mathbf{G}^{<}\right]_{kl},
\end{aligned}
\label{eqn:14}    
\end{equation}
where $D_{ijkl}$ is an appropriate tensor that comprises the spatial correlation between the impurity scattering potentials \cite{camsari2020}. The quantities $\left[\mathbf{G}^{r}\right]_{kl}$, $\left[\mathbf{G}^{<}\right]_{kl}$ represent the retarded Green's function and the lesser Green's function in the matrix representation respectively. For pure dephasing interactions, this tensor has the form 
\begin{equation}
    D_{ijkl} = D_P \delta_{ik}\delta_{jl}
\end{equation}
where $D_P$ is a tunable parameter that controls the strength of the interactions and $\delta_{ij}$ is the Kronecker delta function. For momentum dephasing, the corresponding tensor for these interactions is given by
\begin{equation}
    D_{ijkl} = D_M \delta_{ij}\delta_{ik}\delta_{jl},
\label{eqn:15}    
\end{equation}
where $D_M$ is the corresponding tunable parameter.\\
\indent In the presence of such interactions, the observed tunneling time profile is altered as shown in Fig \ref{fig:6}. Note that both these interactions preserve the spin of the electron and therefore do not affect the measurement mechanism of the setup, which relates to the pointer movement. Thus, all the deviations from coherent tunneling time as observed in Fig \ref{fig:6} indicate disturbances in the actual tunneling process. We note that in the presence of pure phase relaxing interactions, there is an observed broadening in the tunnel time profile. It is also noted that increasing the strength of interactions decreases the tunnel time for electrons with energies close to the height of the barrier. In the presence of momentum relaxing interactions, however, while this effect is less pronounced, a shift in the peak of the function can be observed towards lower energies. This seems to suggest that introducing momentum dephasing interactions to the system lowers the perceived height of the barrier for the electron, consistent with the band-tail effects \cite{Basak_2021} that occur due to momentum relaxation processes. \\
\indent We now study the effect of spin relaxation interactions or equivalently spin dephasing \cite{Yanik} which can be included via
\begin{equation}
\begin{aligned}
\left[\boldsymbol{\Sigma}^{r}_{s}\right]_{ij}&=D_{S}\left(\boldsymbol{\sigma}_{x} \mathbf{G}^{r}_{i, j} \boldsymbol{\sigma}_{x}+\boldsymbol{\sigma}_{y} \mathbf{G}_{i, j}^{r} \boldsymbol{\sigma}_{y}+\boldsymbol{\sigma}_{z} \mathbf{G}_{i, j}^{r} \boldsymbol{\sigma}_{z}\right)\\
\left[\boldsymbol{\Sigma}^{<}_{s}\right]_{ij}&=D_{S}\left(\boldsymbol{\sigma}_{x} \mathbf{G}_{i, j}^{<} \boldsymbol{\sigma}_{x}+\boldsymbol{\sigma}_{y} \mathbf{G}_{i, j}^{<} \boldsymbol{\sigma}_{y}+\boldsymbol{\sigma}_{z} \mathbf{G}_{i, j}^{<} \boldsymbol{\sigma}_{z}\right),
\end{aligned}
\end{equation}
where $\boldsymbol{\sigma}_i$ are the Pauli matrices and $\mathbf{G}^r_{i,j}$ and $\mathbf{G}^<_{i,j}$ correspond to the diagonal $2 \times 2$ sub-blocks of the matrix representation of the retarded Green's function and lesser Green's function respectively.\\
\begin{figure}[h]
    \centering
    \includegraphics[width=0.7\linewidth]{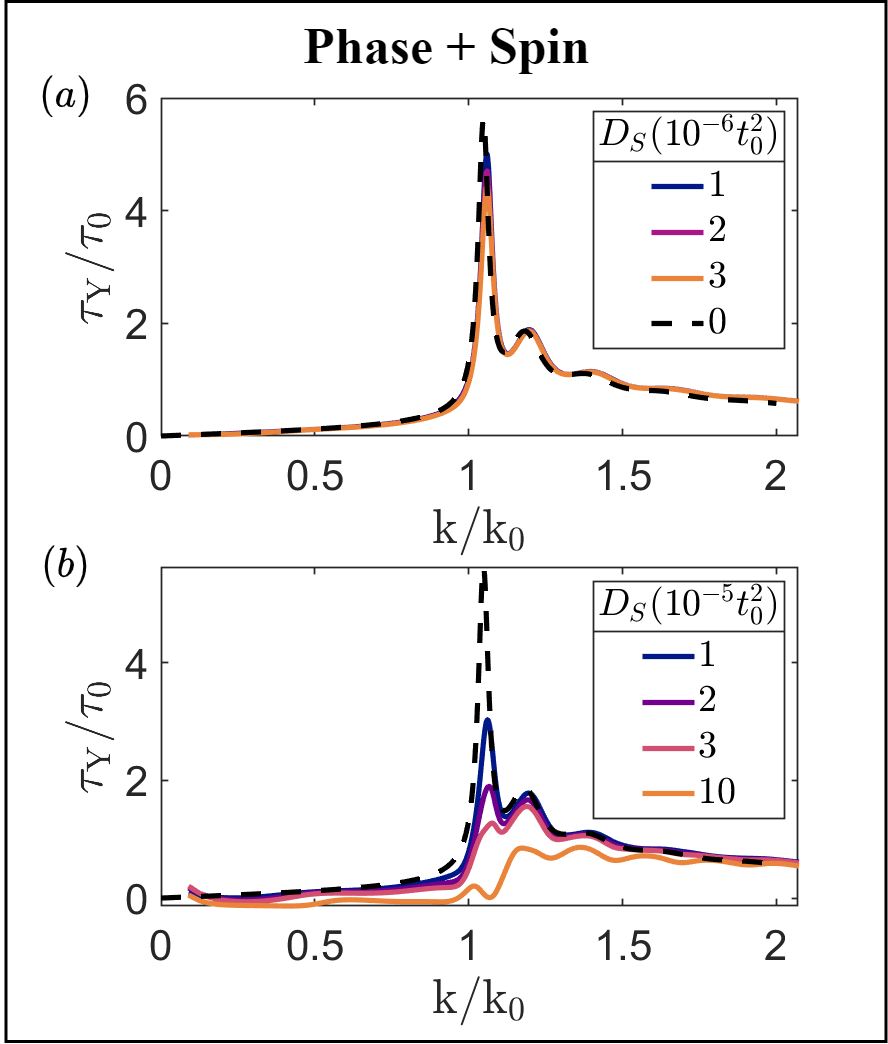}
    \caption{Effect of spin relaxation on the tunnel time profile. (a) For very small values of dephasing parameter we notice some preservation of the tunneling time profile. (b) However, as the dephasing parameter is increased even marginally, we notice a complete break-down of the tunnel time profile.  }
    \label{fig:sdeph}
\end{figure}
This form of the self-energy matrix serves the purpose of re-injecting electrons of opposite spin into the channel thereby relaxing spin. The observed tunnel time in this scenario is altered as shown in Fig \ref{fig:sdeph}. Note that even at very small values of dephasing, the tunnel time signal is completely lost. This agrees with the fact that introducing spin relaxing processes in the system destroys the measurement setup by randomizing the ``pointer apparatus'' itself. Thus we note that the time keeping mechanism for tunnel time and the associated weak values break down in the presence of spin-dephasing, while it remains intact, albeit measuring an altered tunnel time when subject to moderate phase and momentum relaxation. \\
{\it{Conclusion:}}
  In this Letter, we proposed a solid-state implementation of the Larmor clock that exploits tunnel magnetoresistance to distill information on how long the itinerant spins take to traverse a barrier embedded in it. In the coherent transport limit, our analysis provided a direct mapping between the magnetoresistance signals and the tunneling times, thereby aligning with the well-known interpretation of the tunneling time as a quantum weak value. By means of an engineered pre-selection in one of the ferromagnetic contacts, we also elucidated how one can make the measurement ``weak'' by minimizing the back-action, while keeping the tunneling time unchanged. We then analyzed the resulting interpretations of the tunneling time and the measurement back action in the presence of phase breaking effects  \cite{DANIELEWICZ1984239,Datta,PhysRevB.75.081301,7571106,doi:10.1063/1.5023159,doi:10.1063/1.5044254,PhysRevApplied.8.064014,camsari2020,Duse_2021}intrinsic to solid state systems. It is clearly demonstrated that, while the time-keeping aspect of the Larmor clock is reasonably undeterred due to momentum and phase relaxation processes, it degrades significantly in the presence of spin-dephasing. We believe that the ideas presented here can potentially open up an fertile solid state spintronics platform to encompass emerging ideas in quantum technology such as quantum weak values and its applications, that are currently exclusive to quantum optics and cold atoms. While the setup we describe provides a basic realization of a spintronic B\"uttiker clock consistent with the interpretations of Steinberg \cite{Steinberg_1995}, it is left further to look into the thermodynamic aspects of quantum timekeeping via a serious analysis of pointer tick accuracy and efficiency \cite{erker2017,Manoj,Laird}. Furthermore, mesoscopic quantum Hall setups with quantum point contacts also possess realizable configurations for delving deep into these aspects discussed here. Furthermore, the interaction with nuclear spins via the hyperfine interaction can offer new insights into continuous weak measurements \cite{Aniket,Fauzi_1,Fauzi_2}. Most importantly, the general problem of space-time distance estimation in quantum systems is still a matter of intense pursuit and conceptual advancement, where relativistic quantum time dilation may also be possible \cite{Alex}, and suitable test beds may be built featuring quantum materials.\\
{\it{Acknowledgements:}} We wish to acknowledge Supriyo Datta, Sai Vinjanampathy and Ashwin Tulapurkar for useful discussions. The research and development work undertaken in the project under the Visvesvaraya Ph.D Scheme of the Ministry of Electronics and Information Technology (MEITY), Government of India, is implemented by Digital India Corporation (formerly Media Lab Asia). This work is also supported by the Science and Engineering Research Board (SERB), Government of India, Grant No. STR/2019/000030, the Ministry of Human Resource Development (MHRD), Government of India, Grant No. STARS/APR2019/NS/226/FS under the STARS scheme. 

\bibliography{main}

\onecolumngrid 
\appendix

\section*{Supplementary Material}
\renewcommand{\theequation}{S\arabic{equation}}
\renewcommand{\thefigure}{S\arabic{figure}}
\setcounter{equation}{0}
\setcounter{figure}{0}
\subsection{The Keldysh non-equilibrium Green's function technique}
The Keldysh non-equilibrium Green's function (NEGF) method can be used to set up a systematic framework to evaluate the required currents and other quantities. In our formulation of the Keldysh NEGF, the device is connected with multiple leads. First, we consider a channel described by a standard tight-binding Hamiltonian of the form 
\begin{equation}
H = E_0 -t_0  \sum_{i,\sigma} (c_{i\sigma}^\dagger c_{i+1,\sigma}+ \ h.c.),
\end{equation}
where $i \in \{1, N\}$ is the lattice index, $\sigma \in \{-1, 1\}$ is the spin index, and h.c., stands for the hermitian conjugate. In the matrix form, this is given by a $2N \times 2N$ matrix with $E_0 \cdot I_{2\times 2}$ on the $2\times 2$ block diagonals and $-t_0 I_{2\times 2}$ on the $2\times2$ upper and lower off-diagonals.\\
\indent We then add a barrier, with an enclosed magnetic field in the $\hat{z}$ direction, to the channel which has the form
\begin{equation}
U(y) = (V_BI - \frac{V_Z \sigma_z}{2}) \left(\Theta\left(y - \frac{N}{2}+\frac{L}{2}\right) - \Theta\left(y - \frac{N}{2}-\frac{L}{2}\right)\right),
\end{equation}
where $V_B$ is the barrier potential, $I$ is the $2\times2$ identity matrix, $V_Z$ is the Zeeman splitting energy in the $\hat{z}$ direction, $\sigma_z$ is the Pauli matrix in $\hat{z}$, $\Theta(x)$ is the Heaviside step function, $L$ is the length of the barrier, and $N$ is the length of the channel. The matrix form of the above potential is simply given by
\begin{equation}
    U_{ij} = \begin{cases}(V_B I_{2 \times 2} - V_Z \sigma_{z}/2)\delta_{ij} & \mathrm{for} \ N/2-L/2 \leq i \leq N/2-L/2\\
    0 & \mathrm{otherwise}
    \end{cases}
    \label{eqn:barrier}
\end{equation}
where $U_{ij}$ represents the  $2\times2$ block at location $(i.j)$ in the matrix.\\
\indent Now from the device Hamiltonian, $H$, and using \eqref{eqn:barrier}, \eqref{eqn:unpolarized} and \eqref{eqn:polarized}, the retarded Green's function $G^r$ can be obtained as
\begin{eqnarray}
G^r =& [EI-H-U-\Sigma^r_c-\Sigma^r_s]^{-1}, \label{eqn:RetardedGreens-function}
\end{eqnarray}
where $\Sigma^r_c = \sum_{\alpha} \Sigma^r_{\alpha}$ is the sum of all the contact retarded self energies associated with contacts labelled $\alpha$, where $\alpha = FM1, FM2, L, R$ and $\Sigma^r_s$ is the scattering self-energy used to model scattering interactions.\\
\begin{figure}
    \centering
    \includegraphics[width=\textwidth]{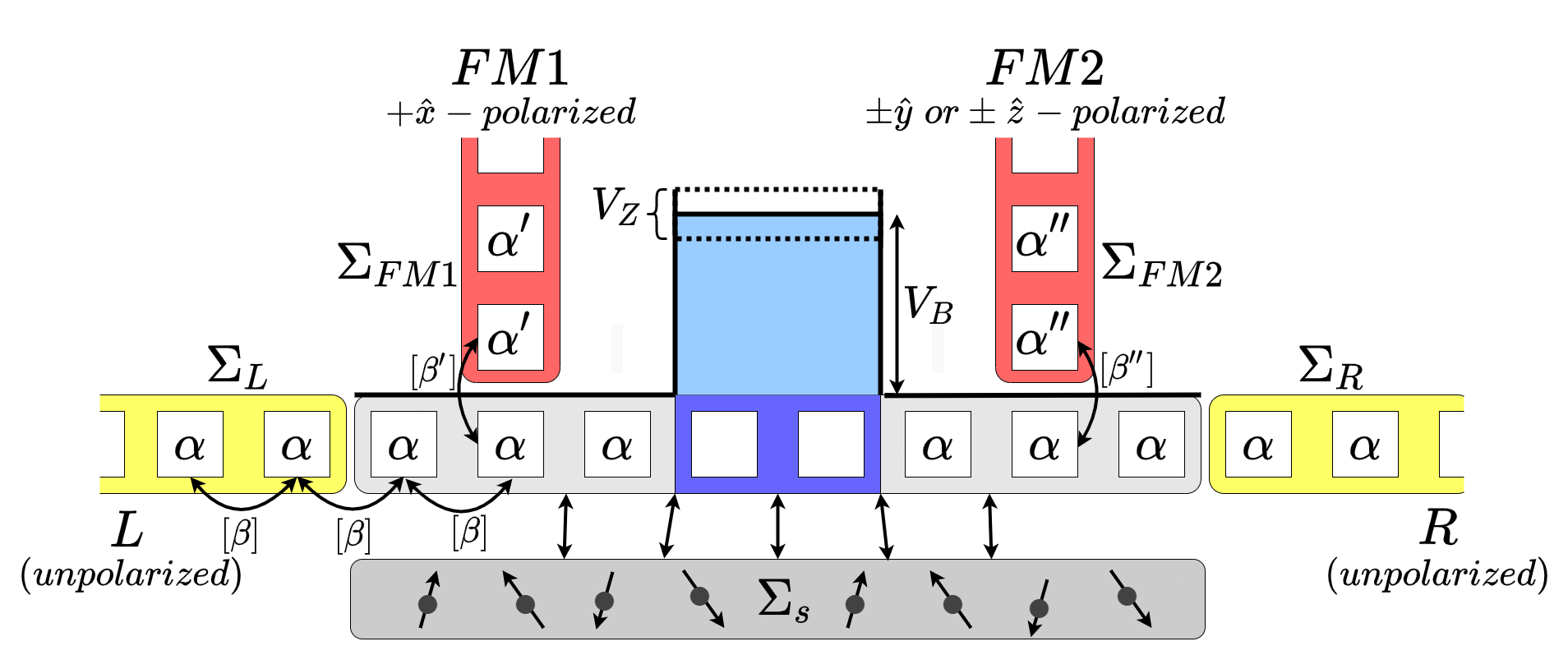}
    \caption{Schematic of the lattice structure of the setup along with a depiction of the contact self energies as well as the dephasing self-energy as used in the NEGF method. $\alpha$, $\beta$ are the on-site and hopping elements in the channel Hamiltonian. Unpolarized contacts (yellow) and polarized contacts (red) are accounted for via the respective self energies. The scatterers that are responsible for dephasing are represented via another bath and are accounted for via an additional self energy.}
    \label{fig:NEGF}
\end{figure}
For an unpolarized contact, the retarded self energy $\Sigma^r_{\alpha}$ attached to a point $i$ in the channel has only two non-zero elements in its $2N\times 2N$ matrix form, given by $[\Sigma^r_{\alpha}]_{i,i} = -t_0e^{ika}I_{2\times 2}$. Thus, the retarded self energy matrices of the two unpolarized contacts in our channel, $\Sigma^r_L$ and $\Sigma^r_R$ are given by:
\begin{equation}
\begin{aligned}
    \left[\Sigma^r_L\right]_{i.j} &= -t_0e^{ika}I_{2\times 2}\delta_{1,i}\delta_{1,j}\\
    [\Sigma^r_R]_{i.j} &= -t_0e^{ika}I_{2\times 2}\delta_{N,i}\delta_{N,j}.
    \label{eqn:unpolarized}
\end{aligned}
\end{equation}
For a perfectly polarized ferromagnetic contact polarized in the $\hat{p_{\alpha}}$ direction, and located at a point $i$ in the channel, the self energy matrix has a block diagonal form given by
\begin{equation}
[\Sigma^r_{\alpha}]_{ii} = -t_0 e^{ika}(I_{2\times2} + \mathbf{\hat{p_{\alpha}}\cdot \boldsymbol{\sigma}})/2,
\end{equation}
where $\boldsymbol{\sigma}$ are the Pauli matrices.\\
Thus, the self energy matrices of the two FM polarized contacts in our channel, $\Sigma^r_{FM1}$ and $\Sigma^r_{FM2}$ are given by:
\begin{equation}
\begin{aligned}
    \left[\Sigma^r_{FM1}\right]_{i.j} &= -\frac{t_0 e^{ika}}{2}(I_{2\times2} +\boldsymbol{\sigma}_x)\delta_{(N-L)/4,i} \ \delta_{(N-L)/4,j}\\
    [\Sigma^r_{FM2}]_{i.j} &= \begin{cases} -\frac{t_0 e^{ika}}{2}(I_{2\times2} \pm \boldsymbol{\sigma}_y)\delta_{(3N+L)/4,i} \ \delta_{(3N+L)/4,j} & \text{while measuring $\tau_Y$}\\
    -\frac{t_0 e^{ika}}{2}(I_{2\times2} \pm \boldsymbol{\sigma}_z)\delta_{(3N+L)/4,i} \ \delta_{(3N+L)/4,j} & \text{while measuring $\tau_Z$}.
    \end{cases}
\end{aligned}
\label{eqn:polarized}
\end{equation}
Note that the $FM1$ contact is located at position $(N-L)/4$ and the $FM2$ contact is located at position $(3N+L)/4$.\\
Now, the lesser self energies of the contacts are given by
\begin{equation}
    \Sigma^{<}_{\alpha} = i \Gamma_{\alpha} f_{\alpha},
    \label{eqn:inscattering}
\end{equation} 
where $\Gamma_{\alpha}$ is simply the broadening function given by the imaginary part of the self energy, such that, $\Gamma_{\alpha} = i[\Sigma^r_{\alpha} - \Sigma^a_{\alpha}]$, and $f_m$ is the corresponding occupation factor.\\
Then, the lesser Green's function $G^<$ at a particular energy is given by:
\begin{equation}
    G^< = G^r[\Sigma^{<}_c + \Sigma^{<}_s]G^a \label{eqn:Greens-function}
\end{equation}
where $G^a=[G^r]^{\dagger}$ is the advanced Green's function, $\Sigma^{<}_c = \sum_{\alpha} \Sigma^{<}_{\alpha}$ is the sum of all contact lesser self energies and  $\Sigma_s^{<}$ is the lesser self-energy arising from the dephasing interactions. $G^n = -i G^{<}$ represents the electron density (times $2\pi$) inside the channel.\\
The spectral function $A$ is obtained as,
\begin{eqnarray}
A = i[G^r - G^a] = G^r[\Gamma_c + \Gamma_s]G^a
\label{eqn:spectral-function}
\end{eqnarray}
The diagonal elements of the spectral function are related to the local density of states (LDOS) at the corresponding lattice point in the channel. We are now left with defining the scattering self energies due to dephasing processes considered in this work.\\
{\it{Dephasing Self-energies:}} To account for scattering in the contacts, we introduce a self-energy matrix for the various dephasing processes. We consider impurities with localised potentials $U_s(i)$ in the channel and use their correlator $\bar{D}(i,j) = \braket{U_s(i)}{U^*_s(j)}$ to calculate the self-energy of interaction processes. This facilitates a smooth transition from the ballistic regime to the diffusive regime.\\
The self-energy for the momentum dephasing process is given by,
\begin{eqnarray}
\Sigma^r_s(i,j) =& \bar{D}(i,j)G^r(i,j) \\
\Sigma^{<}_s(i,j) =& \bar{D}(i,j)G^<(i,j)
\label{eqn:dephasing-self-energies}
\end{eqnarray}
where $G^<(i,j)$  is the lesser Green's function and $\bar{D}(i,j)$ is given by,
\begin{eqnarray}
\bar{D}(i,j) =& \braket{U_s(i)}{U^*_s(j)} \\
\bar{D}(i,j) =& D_m \delta_{ij}
\label{eqn:momentum-dephasing}
\end{eqnarray}
This model discards the off-diagonal elements of the Green’s function, thus relaxing both the phase and momentum of quasiparticles in the channel. The quantity $D_m$ is the dephasing parameter
which represents the magnitude squared of the fluctuating scattering potentials. This parameter can be modulated so that by gradually increasing it, one can transition from the coherent ballistic limit to the diffusive limit.\\
Similarly, the self-energy for pure phase dephasing process is given by,
\begin{eqnarray}
\Sigma^r_s =& \bar{D}_p G^r \\
\Sigma^{<}_s =& \bar{D}_p G^<
\label{eqn:pure-dephasing-self-energies}
\end{eqnarray}
where $\bar{D}_p$ is again the dephasing parameter that controls the magnitude of interactions. Here, the entire Green's function is preserved as the self-energy matrix and relaxes only the phase of the quasiparticles.\\
Spin-flip interactions can be added to the channel via the introduction of a corresponding self energy of the form:
\begin{equation}
\begin{aligned}
\left[\boldsymbol{\Sigma}^{r}_{s}\right]_{ij}&=D_{S}\left(\boldsymbol{\sigma}_{x} \mathbf{G}^{r}_{i, j} \boldsymbol{\sigma}_{x}+\boldsymbol{\sigma}_{y} \mathbf{G}_{i, j}^{r} \boldsymbol{\sigma}_{y}+\boldsymbol{\sigma}_{z} \mathbf{G}_{i, j}^{r} \boldsymbol{\sigma}_{z}\right)\\
\left[\boldsymbol{\Sigma}^{<}_{s}\right]_{ij}&=D_{S}\left(\boldsymbol{\sigma}_{x} \mathbf{G}_{i, j}^{<} \boldsymbol{\sigma}_{x}+\boldsymbol{\sigma}_{y} \mathbf{G}_{i, j}^{<} \boldsymbol{\sigma}_{y}+\boldsymbol{\sigma}_{z} \mathbf{G}_{i, j}^{<} \boldsymbol{\sigma}_{z}\right).
\end{aligned}
\end{equation}
The effect of this dephasing mechanism is to reinject
an electron with an opposite spin back to the channel that relaxes spin.\\
\textit{Current Operator:} The NEGF formalism provides us a clear cut current operator that can be used to calculate all kinds of currents through a contact $\alpha$, given by:
\begin{equation}
\mathbf{I}^{\alpha}_{\mathrm{op}}= \frac{1}{h}\left(\left[\mathbf{\Sigma^r_{\alpha}} \mathbf{G}^{<}-\mathbf{G}^{<} \mathbf{\Sigma}^{a}_{\alpha}\right]+\left[\mathbf{\Sigma}^{<}_{\alpha} \mathbf{G}^{a}-\mathbf{G}^{r} \mathbf{\Sigma}^{<}_{\alpha}\right]\right)
\end{equation}
where the current of a particular quantity $X$ is given by $I^{\alpha}_X = Trace(I^{\alpha}_{op}X_{op})$
To find the charge current through the contact, we then simply need to find $I_{\alpha} = Trace(I^{\alpha}_{op})$ since the charge operator is the identity matrix (times e). Then, the charge current per unit energy through a particular contact $m$, is given by 
\begin{equation}
    \tilde{I}_{\alpha}=\frac{-ie}{h} \operatorname{Trace}\left[\Sigma_{\alpha}^{<} A- \Gamma_{\alpha} {G}^{<}\right]
    \label{eqn:current}
\end{equation}
In our setup, we measure the currents through the right unpolarized contact (R) as well as the right ferromagnetic contact (FM2), both of which are grounded.\\
Using \eqref{eqn:Greens-function} solved self consistently with the equations for the retarded Green’s functions given in Eqn \eqref{eqn:RetardedGreens-function}
and self energies given in  \eqref{eqn:unpolarized},\eqref{eqn:polarized},\eqref{eqn:momentum-dephasing}, and \eqref{eqn:pure-dephasing-self-energies}, we next obtain the currents given by \eqref{eqn:current}. We then calculate the spin polarization in the channel using the currents through the respective contacts.
\subsection{Transport signals for the tunneling time and the back action}
In order to measure the dwell time of the electron, the average in-plane precession of the tunneled electron needs to be measured. In other the words, a weak value measurement of the $\sigma_Y$ operator is required. To do this, we consider the right polarized contact in the $\pm \hat{y}$ directions. To understand this measurement, consider two spin channels in the device , carrying electrons of spin polarization $+\hat{y}$ and   $-\hat{y}$. Electrons that are in a superposition of these states travel simultaneously through both channels. The presence of a $+\hat{y}$ polarized ferromagnetic contact then acts as a fork in the channel; with an electron in the $+\hat{y}$ channel draining into either the ferromagnetic $+\hat{y}$ contact, or the unpolarized contact. On the other hand, an electron in the $-\hat{y}$ channel can only drain into the unpolarized contact. In addition to this, the Zeeman field in the barrier, being in the $+\hat{z}$ direction, couples the wave-functions in the two channels.
\begin{figure}
    \includegraphics[width = \textwidth]{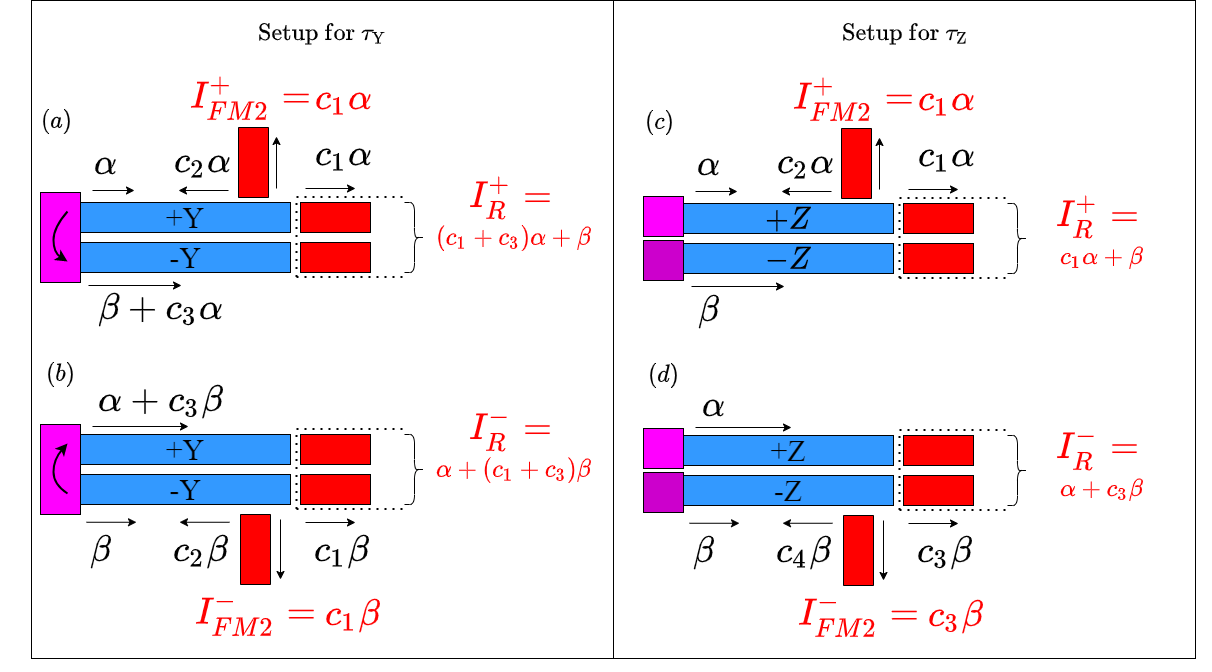}
    \caption{(a), (b): Schematic of the channel "downstream" of the barrier for the right polarized contact in $+\hat{y}$ and $-\hat{y}$ directions. Electrons that have tunneled through the barrier (pink) move along the two spin channels to the right where they are collected by the contacts (red). (c), (d): Schematic of the channel "downstream" of the barrier for the right polarized contact in $+\hat{z}$ and $-\hat{z}$ directions. Note that unlike in (a) and (b), the barrier (pink) is now of different  heights for the two spin channels}
    \label{fig:2}
\end{figure}
Now, let the currents through the two channels, in the absence of any forks be given by $\alpha$ and $\beta$ respectively for the $\pm \hat{y}$ channels. The presence of a fork in the $+\hat{y}$ channel then modifies the current through each of the forks to be $c_1 \alpha$ where $c_1$ is some constant parameter (since the forks are identical). Note that this is independent of $\beta$ since the wave-function in the $-\hat{y}$ channel does not affect the current through the $+\hat{y}$ channel. However, the opposite is not true. The reflected wave function from the fork in the $+\hat{y}$ channel is coupled to the $-\hat{y}$ channel through the $+\hat{z}$ Zeeman field in the barrier. Thus the current through this channel is given by $\beta+ c_3 \alpha$ where $c_3$ is some other constant parameter.

Then, in the presence of the ferromagnetic $+\hat{y}$ polarized contact, the currents through the two contacts are now given by $I^{+}_{FM2}=c_1 \alpha$ (for the $+\hat{y}$ polarized contact) and $I^{+}_{R}=(c_1+c_3)\alpha + \beta$ (for the unpolarized contact)

For the ferromagnetic contact polarized in the $-\hat{y}$ direction, the fork is now in the $-\hat{y}$ channel. Then, the current through each of the forks in the $-\hat{y}$ channel is $c_1\beta$ and the current through the $+\hat{y}$ channel is $\alpha+c_3 \beta$. It is important to note that the constant parameters $c_1, \ c_3$ remain the same in both these cases. This is due to the symmetry of the $\pm \hat{y}$ directions w.r.t the $+\hat{z}$ Zeeman field. Since the $\pm \hat{y}$ directions are indistinguishable w.r.t. the $+\hat{z}$ Zeeman field, the systems are also indistinguishable in the two cases and thus, their constant parameters remain the same.
Then, in the presence of the ferromagnetic $-\hat{y}$ polarized contact, the currents through the two contacts are now given by $I^{-}_{FM2}=c_1 \beta$ (for the $-\hat{y}$ polarized contact) and $I^{-}_{R}=(c_1+c_3)\beta + \alpha$ (for the unpolarized contact)

To measure the polarization of the electron, we will now define a physically observable quantity called "signature" of the electron, given by 
\begin{equation}
D_Y = \frac{I^{+}_{FM2} - I^{-}_{FM2}}{I^{+}_{FM2} + I^{-}_{FM2}},
\end{equation}
where $I^{\pm}_{FM2}$ is the current through the ferromagnetic contact while it is in the $\pm \hat{y}$ orientation respectively.
  
This "signature", is a measure of the spin polarisation of the tunnelled electron. In our simulations, it will play the same role as the quantity $\left<S_Y\right>/(\hbar/2)$ defined by B{\"u}ttiker. Note that this is not the spin expectation value of the wave-function in a particular direction. It is the Y-spin expectation value of the post-selected part of the wave-function.

It is easy to see that,
\begin{equation}
D_Y = \frac{I^{+}_{FM2} - I^{-}_{FM2}}{I^{+}_{FM2} + I^{-}_{FM2}} = \frac{(\alpha - \beta)}{(\alpha + \beta)},
\end{equation}
which gives us back the polarisation of spin in the $\hat{y}$ direction the channel \\
Then the weak value of the $\hat{y}-spin$ of the electron is given by
\begin{equation}
\left<S_Y\right> = (\hbar/2) \frac{I^{+}_{FM2} - I^{-}_{FM2}}{I^{+}_{FM2} + I^{-}_{FM2}} = -(\hbar/2)\omega_L \tau_Y.
\end{equation}
Thus we have
\begin{equation}
\tau_Y = -\frac{1}{\omega_L} \frac{I^{+}_{FM2} - I^{-}_{FM2}}{I^{+}_{FM2} + I^{-}_{FM2}}.
\end{equation}
To measure the measurement back-action in this setup, the out-of plane alignment of the spin polarization is to be measured. This is equivalent to a weak value measurement of the $\sigma_Z$ operator. Thus, in this setup, the right ferromagnetic contact is polarized in the $\pm \hat{z}$ directions. 

Since we have to measure the polarisation in the $\hat{z}$ direction, we now consider spin channels carrying electrons of spin polarization $+\hat{z}$ and $-\hat{z}$. The primary difference to note here is that in this setup, the channels are not coupled to each other since the Zeeman field is in the $\hat{z}$ direction. As a result, the two setups corresponding to $\pm \hat{z}$ polarized ferromagnetic contact are non-identical to each other.

Once again, let the currents through the two channels, in the absence of any forks be given by $\alpha$ and $\beta$ respectively for the $\pm \hat{z}$ channels. Just as before, the presence of a fork in the $+\hat{z}$ channel then modifies the current through each of the forks to be $c_1 \alpha$ where $c_1$ is some constant parameter (since the forks are identical). However, the current through the $-\hat{z}$ channel is unmodified in this case and simply $\beta$.

Thus, in the presence of the ferromagnetic $+\hat{z}$ polarized contact, the currents through the two contacts are now given by $I^{+}_{FM2}=c_1 \alpha$ (for the $+\hat{z}$ polarized contact) and $I^{+}_{R}=c_1\alpha + \beta$ (for the unpolarized contact).

Similarly, in the presence of the ferromagnetic $-\hat{z}$ polarized contact, the currents through the two contacts are now given by $I^{-}_{FM2}=c_3 \beta$ (for the $+\hat{z}$ polarized contact) and $I^{-}_{R}=c_3\beta + \alpha$ (for the unpolarized contact).

Note that in this case, the constant parameters are not the same since the channels are no longer identical. However, we still need polarization of the form $(\alpha-\beta)/(\alpha+\beta)$. This can be realized with a different choice of "signature" given by:
\begin{equation}
D_Z= \frac{(I_R^{+}-I_{FM2}^{+}) - (I_R^{-}-I_{FM2}^{-})}{(I_R^{+}-I_{FM2}^{+}) + (I_R^{-}-I_{FM2}^{-})},
\end{equation}
where $I^{\pm}_{FM2}$ are, as in the previous case, the current through the ferromagnetic contact while it is in the $\pm \hat{z}$ orientation respectively. $I_{R}^{\pm}$ is the current through the right unpolarized contact when the ferromagnetic contact is in the $\pm \hat{z}$ orientation respectively.

It is easy to see that $I_R^{+}-I_{FM2}^{+} = \beta$ and similarly, $I_R^{-}-I_{FM2}^{-} = \alpha$.  Thus, we have,
\begin{equation}
D_Z=-\frac{(I_R^{+}-I_{FM2}^{+}) - (I_R^{-}-I_{FM2}^{-})}{(I_R^{+}-I_{FM2}^{+}) + (I_R^{-}-I_{FM2}^{-})} = \frac{(\alpha-\beta)}{( \alpha+\beta)}.
\end{equation}
Then the weak value of the $\hat{z}-spin$ of the electron is given by
\begin{equation}
\left<S_Z\right> = -(\hbar/2) \frac{(I_R^{+}-I_{FM2}^{+}) - (I_R^{-}-I_{FM2}^{-})}{(I_R^{+}-I_{FM2}^{+}) + (I_R^{-}-I_{FM2}^{-})} = (\hbar/2)\omega_L \tau_Z.
\end{equation}
We then have
\begin{equation}
\tau_Z = -\frac{1}{\omega_L} \frac{(I_R^{+}-I_{FM2}^{+}) - (I_R^{-}-I_{FM2}^{-})}{(I_R^{+}-I_{FM2}^{+}) + (I_R^{-}-I_{FM2}^{-})}.
\end{equation}
It is also important to note that this choice of "signature" only holds true in the tunneling regime. When the particle is no longer tunneling, it is possible for the contact reflections to traverse back through the barrier and interact with the +X contact which in turn couples the $\pm Z$ channels with each other. This violates our initial assumption that the current in the unforked channel is unaffected and thus distorts the results. In the tunneling regime, these reflections (which have to tunnel back through the barrier and then through it again)  are of very low magnitude to cause any significant distortions in the result. \\
\subsection{Weak values}
When one tries to measure a variable attached to an operator, say $\hat{A}$, using a ``pointer'' generated via an operator, say $\hat{P}$, the generic interaction for the measurement within the von-Neumann framework \cite{Jordan_PRL,Boaz} is given by
\begin{equation}
\hat{H}_{int} = - g(t) \hat{P} \otimes \hat{A}.
\end{equation}
Here, $g(t)$ is a compact supported function in the duration of the measurement such that $\gamma = \int g(t) dt$ represents a small coupling parameter that characterizes this interaction. Based on the pre-selection and post-selection, we can then characterize the measurement of an eigenvalue $a_n$ of the operator $\hat{A}$ via the movement of the conjugate $\hat{Q}$ of the pointer variable $\hat{P}$, as derived quite explicitly in Ref. \cite{Sudarshan}. \\
\indent We need to measure the expectation of A, given the initial and final states $\ket{i}$ and $\ket{f}$. This is given as
$$\left<A\right>_{fi} = \frac{\braket{f|A|}{i}}{\braket{f}{i}}$$.
Let us assume that the pointer was initially in the state,
$$\ket{\psi}_i = \exp{(-Q^2/4\sigma^2)}$$
where $\sigma^2$ is the variance of $Q$. After the measurement, this then transforms to 
$$
\begin{aligned}
\ket{\psi}_f &= \exp{\left(\frac{-(Q-\left<A \right>_{fi})^2}{4\sigma^2}\right)}
=\exp{\left(\frac{-(Q-\Re\left<A \right>_{fi} -i\Im\left<A \right>_{fi})^2}{4\sigma^2}\right)}\\
&=\exp\left(\frac{-(Q-\Re\left<A \right>_{fi})^2}{4\sigma^2}\right)
\exp\left(\frac{-(\Im\left<A \right>_{fi})^2}{4\sigma^2}\right)
\exp\left( \frac{-i\Im\left<A \right>_{fi}(Q-\Re\left<A \right>_{fi})}{2\sigma^2}   \right).
\end{aligned}
$$
From the above, we see that the change in pointer position is given by
\begin{equation}
    \Delta Q= \Re\left<A \right>_{fi},
    \label{eqn:Pointer-position}
\end{equation}
whereas the change in pointer momentum is the associated phase of the form $exp(i p\cdot q)$ which means the change in pointer momentum,
\begin{equation}
  \Delta P = \Im\left<A \right>_{fi}/2\sigma^2, 
  \label{eqn:Pointer-momentum}
\end{equation}
where $\Delta Q$ and $\Delta P$ are the "weak" values. However, this does not mean the measurement is weak. It becomes weaker when $Q$ is more uncertain and thus, $\sigma$ increases as will be shown in the next part.
\subsection{Weak measurements in our setup}
In our setup, the pointer is $\phi$ and therefore, the conjugate momentum is $S_Z$. The operator, whose expectation value needs to be measured, as shown by Steinberg, is $U(y) = 1/2(\Theta(y+L/2)-\Theta(y-L/2))$, where $\Theta(y)$ is the Heaviside step function. Here the barrier is from $y=-L/2$ to $y=L/2$. To couple this operator to the pointer as in \eqref{eqn:1}, we put a magnetic field inside the barrier, in the Z direction. Thus, the Hamiltonian becomes,
\begin{equation}
    \hat{H} = -\gamma B_z \hat{S}_z \hat{U}(\hat{y}).
\end{equation}
Then, as shown in  \eqref{eqn:Pointer-position} and \eqref{eqn:Pointer-momentum}, the change in pointer position,
\begin{equation}
    \Delta \phi = \omega_L \tau_Y = k\Re \left<U(y)\right>_{fi},
    \label{eqn:thetaxy}
\end{equation}
where $\omega_L$ is the Larmor frequency. The change in pointer momentum,
\begin{equation}
\begin{aligned}
    \Delta S_Z = \omega_L \tau_Z &= k\Im \left<U(y)\right>_{fi}/(2 \ Var(\phi))\\
    &=k\Im \left<U(y)\right>_{fi}*Var(S_Z)/2.
    \label{eqn:delta-sz}
\end{aligned}
\end{equation}
Since our initial state is in the $\hat{x}$ polarized direction, $\left<S_Z \right>_i = 0 \implies \Delta S_Z= \left<S_Z \right>_f - \left<S_Z \right>_i= \left<S_Z \right>_f$.\\
Thus, the pointer momentum measured by our device is indeed the correct and required change in pointer momentum. However, this is not the case for $\phi$. We are capable of measuring only the expectation value of spins of the electron and not the angle in x-y plane. Specifically, this angle is the phase difference between the two components of the spinor. Consider a spinor of the form
\begin{equation}
\begin{aligned}
    \ket{\psi} &= \begin{bmatrix}
    \cos \theta/2 \\ \sin\theta/2 \ e^{i\phi}
    \end{bmatrix}\\
    \text{Then, } \braket{\psi}{S_Y|\psi} &= 2 \sin(\theta/2) \cos (\theta/2) \sin \phi =\sin \phi \sin \theta.
\end{aligned}
\label{eqn:spinor}
\end{equation}
If our spinor is initially in the state $\theta = \pi/2$, we have,
$$\braket{\psi}{S_Y|\psi} = \sin \phi \approx \phi \text{ when $\phi \rightarrow 0$}$$. 
Thus we can obtain the change in $\phi$ via the measurement of $S_Y$ only when $\phi$ is very small. Now we need to find $\phi$ in the case where $\theta \neq \pi/2$. Note that 
\begin{equation}
    \braket{\psi}{S_Z|\psi} = \cos^2\theta/2 - \sin^2 \theta/2.
\end{equation}
Then the variance in $S_Z$,
\begin{equation}
    Var(S_Z) = 1 - \left<S_Z \right>^2 = 1 - \cos^2 \theta = \sin^2\theta.
    \label{eqn:varsz}
\end{equation}
Combining \ref{eqn:spinor} and \ref{eqn:varsz}, we have,
\begin{equation}
    \frac{\braket{\psi}{S_Y|\psi}}{\sqrt{Var(S_Z)}}= \sin \phi \approx \phi \text{ when $\phi \rightarrow 0$}.
    \label{eqn:constant-theta}
\end{equation}
From \ref{eqn:thetaxy}, we know that $\Delta \phi$ should remain constant even when $Var(\phi)$ changes. Thus, from \eqref{eqn:constant-theta}, we see that $\frac{\braket{\psi}{S_Y|\psi}}{\sqrt{Var(S_Z)}}$ must remain constant and this is verified using NEGF.\\
Similarly, from \eqref{eqn:delta-sz}, we see that $\Delta S_Z * Var(\phi)$ must remain constant which means, $\Delta S_Z/Var(S_{Z})$ must remain constant, as is quite well verified using our NEGF approach also. 
\\ \\
Thus, we see that the measurement can be made weaker by changing the polarization of the electron to $\phi<\pi/2$. This decreases the uncertainty in $S_Z$ and therefore increases the uncertainty in $\phi$. The measurement back-action, $\Delta S_Z$ is found to decrease proportionally to the decrease in $Var(S_Z)$. The real part of the measurement, $\phi$ remains a constant, but since that cannot be explicitly measured, we show that the measured quantity, $\braket{\psi}{S_Y|\psi}$ decreases proportionally to the square root of $Var(S_Z)$ and that $\phi= \frac{\braket{\psi}{S_Y|\psi}}{\sqrt{Var(S_Z)}}$ is the actual real part of the measurement. 

\end{document}